\documentclass[prd,aps,showpacs,onecolumn]{revtex4}
\usepackage{amssymb}
\usepackage{amsfonts}
\usepackage{amsmath}
\usepackage{graphicx}
\usepackage{bm}

\begin{document}

\title{Solvable rational extensions of the Morse and Kepler-Coulomb
potentials}
\author{Yves Grandati }
\affiliation{Institut de Physique, Equipe BioPhyStat, ICPMB, IF CNRS 2843, Universit\'{e}
Paul Verlaine-Metz, 1 Bd Arago, 57078 Metz, Cedex 3, France}

\begin{abstract}
We show that it is possible to generate an infinite set of solvable rational
extensions from every exceptional first category translationally shape
invariant potential. \bigskip This is made by using Darboux-B\"{a}cklund
transformations based on unphysical regular Riccati-Schr\"{o}dinger
functions which are obtained from specific symmetries associated to the
considered family of potentials.
\end{abstract}

\maketitle

\section{\protect\bigskip}

\section{Introduction}

In the recent years, several notable progresses have been made in the
research and characterization of new closed-form exactly solvable systems in
quantum mechanics \cite%
{gomez,gomez2,gomez3,gomez4,gomez5,gomez6,quesne1,quesne,quesne2,quesne3,odake,sasaki,ho,odake2,sasaki2,dutta,grandati2,grandati3}%
. The obtained systems are regular rational extensions of some
shape-invariant potentials \cite{cooper,Dutt,Gendenshtein} and are
associated to families of exceptional orthogonal polynomials (EOP) built
from the Laguerre or Jacobi classical orthogonal polynomials. In all the
considered cases, the initial potentials belong to the second category (as
defined in \cite{grandati}) of primary translationally shape-invariant
potentials (TSIP):\ the extended potentials of the $J1$ and $J2$ series
(associated to the Jacobi EOP) are obtained from the generic second category
potentials (Darboux-P\"{o}schl-Teller or Scarf of the hyperbolic or
trigonometric types), as for the extended potentials of the $L1,$ $L2$ and $%
L3$ series, they are obtained from the unique exceptional second category
potential which is the isotonic one.

If we except the specific case of the harmonic oscillator which has been
extensively treated \cite%
{shnol',samsonov,tkachuk,gomez,carinena,fellows,grandati2}, the solvable
extensions of first category potentials have been much less studied.
Refering to the classification established in \cite{grandati}, the
exceptional first category primary TSIP are the one-dimensional harmonic
oscillator (HO), the Morse potential and the effective radial Kepler-Coulomb
(ERKC) system, whereas the generic first category primary TSIP include the
trigonometric and hyperbolic Rosen-Morse potentials as well as the Eckardt
potential. A general study of the possible extensions of a large number of
exactly solvable potentials from the point of view of conditionally solvable
potentials has been made by Junker and Roy \cite{junker}. The case of the
Morse potential has been also considered by Gomez-Ullate et al \cite{gomez}
who have determined the algebraic deformations of this system which are
solvable by polynomials.

In \cite{grandati3} we have developped a new approach which allows to
generate an infinite set of regular exactly solvable extensions starting
from every TSIP in a very direct and systematic way without taking recourse
to any ansatz. This approach is based on a generalization of the usual SUSY
partnership built from excited states. The corresponding Darboux-B\"{a}%
cklund Transformations (DBT), which are covariance transformations for the
class of Riccati-Schr\"{o}dinger (RS) equations \cite{grandati}, are based
on regularized RS functions which are obtained by using discrete symmetries
acting on the parameters of the considered family of potentials. Considering
the isotonic oscillator, we have obtained the three infinite sets $L1$, $L2$
and $L3$ of regular rationally solvable extensions of this potential and
have given a simple and transparent proof of the shape-invariance of the
potentials belonging to the $L1$ and $L2$ series. In the present paper we
show that the same approach can be applied to generate infinite towers of
solvable rational extensions from every exceptional first category
potential. As shown in \cite{grandati}, the first category primary TSIP can
be reduced into a harmonic one by a change of the variable which satisfies a
constant coefficient Riccati equation. The exceptional cases correspond to
the cases where this equation degenerates into a linear equation or a
Riccati equation with a double root in the right-hand member, namely the HO,
the Morse and ERKC potentials. In this cases the bound states are
expressible in terms of generalized Laguerre Polynomials (GLP) \cite%
{cooper,Dutt}.

The paper is organized as follows. After recalling briefly the basic
elements of our approach, we test its efficiency on the simple and
exhaustively studied case of one-dimensional HO, retrieving very simply the
results already obtained in \cite%
{shnol',samsonov,tkachuk,gomez,carinena,fellows,grandati2}. In the second
and third parts, we treat successively the Morse and ERKC systems, building
the associated towers of solvable regular extensions and characterizing
their eigenstates. For the Morse potential we recover the algebraic
deformations described by Gomez-Ullate et al \cite{gomez}, the extensions
being not strictly isospectral to the primary potential. For the ERKC
potential we obtain two disctinct regimes with respect to the value of the
"angular momentum" parameter. In the first regime the extensions are
strictly isospectral to the primary potential whereas in the second regime
they are not.

Contrarily to the case of the second category potentials the extensions of
the exceptional first category potentials do not inherit of the shape
invariance properties of the primary potential.

\section{Darboux-B\"{a}cklund Transformations (DBT) and regular extensions}

\subsection{General scheme}

Consider a family of one-dimensional hamiltonians\ indexed by a
multiparameter $a$

\begin{equation*}
\widehat{H}(a)=-d^{2}/dx^{2}+V(x;a),\ a\in \mathbb{R}^{m},\ x\in I\subset 
\mathbb{R}
\end{equation*}

If $\psi _{\lambda }(x;a)$ is an eigenstate of $\widehat{H}(a)$ associated
to the eigenvalue $E_{\lambda }(a)$, then its logarithmic derivative $%
w_{\lambda }(x;a)=-\psi _{\lambda }^{\prime }(x;a)/\psi _{\lambda }(x;a)$,
that we will call a Riccati-Schr\"{o}dinger (RS) function, satisfies a
particular Riccati equation of the following form

\begin{equation}
-w_{\lambda }^{\prime }(x;a)+w_{\lambda }^{2}(x;a)=V(x;a)-E_{\lambda }(a).
\label{edr4}
\end{equation}

Eq(\ref{edr4}) is called the Riccati-Schr\"{o}dinger (RS) equation \cite%
{grandati} for the level $E_{\lambda }(a)$. The RS function $w_{\lambda
}(x;a)$ presents a simple pole at each node of the eigenstates $\psi
_{\lambda }(x;a)$.

It is a well-known fact that the set of general Riccati equations is
invariant under the group $\mathcal{G}$ of smooth $SL(2,\mathbb{R})$-valued
curves $Map(\mathbb{R},SL(2,\mathbb{R}))$ \cite{carinena2,Ramos}. The
particular of Riccati-Schr\"{o}dinger equations is, as for it, preserved by
a specific subset of $\mathcal{G}$. These transformations, called Darboux-B%
\"{a}cklund Transformations (DBT), are build from any solution $w_{\nu
}(x;a) $ of the initial RS equation Eq(\ref{edr4}) as \cite%
{carinena2,Ramos,grandati}

\begin{equation}
w_{\lambda }(x;a)\overset{A\left( w_{\nu }\right) }{\rightarrow }w_{\lambda
}^{\left( \nu \right) }(x;a)=-w_{\nu }(x;a)+\frac{E_{\lambda }(a)-E_{\nu }(a)%
}{w_{\nu }(x;a)-w_{\lambda }(x;a)},  \label{transfoback2}
\end{equation}%
where $E_{\lambda }(a)>E_{\nu }(a)$. Then $w_{\lambda }^{\left( \nu \right)
} $ is a solution of the RS equation:

\begin{equation}
-w_{\lambda }^{\left( \nu \right) \prime }(x;a)+\left( w_{\lambda }^{(\nu
)}(x;a)\right) ^{2}=V^{\left( \nu \right) }(x;a)-E_{\lambda }(a),
\label{eqtransform}
\end{equation}%
with the same energy $E_{\lambda }(a)$ as in Eq(\ref{edr4}) but with a
modified potential

\begin{equation}
V^{\left( \nu \right) }(x;a)=V(x;a)+2w_{\nu }^{\prime }(x;a).
\label{pottrans}
\end{equation}

The corresponding eigenstate of $\widehat{H}^{\left( \nu \right)
}(a)=-d^{2}/dx^{2}+V^{\left( \nu \right) }(x;a)$ can be written

\begin{equation}
\psi _{\lambda }^{\left( \nu \right) }(x;a)=\exp \left( -\int dxw_{\lambda
}^{(\nu )}(x;a)\right) \sim \frac{1}{\sqrt{E_{\lambda }\left( a\right)
-E_{\nu }(a)}}\widehat{A}\left( w_{\nu }\right) \psi _{\lambda }(x;a),
\label{foDBT}
\end{equation}%
where $\widehat{A}\left( a\right) $ is a first order operator given by

\begin{equation}
\widehat{A}\left( w_{\nu }\right) =d/dx+w_{\nu }(x;a).  \label{opA}
\end{equation}

From $V$, the DBT generates a new potential $V^{\left( \nu \right) }$
(quasi) isospectral to the original one and its eigenstates are directly
obtained from those of $V$ via Eq(\ref{foDBT}). If the initial system is
exactly solvable, which is the case of the translationally shape invariant
potentials (TSIP), this scheme allows to build new exactly solvable
potentials.

Nevertheless, in general, $w_{\nu }(x;a)$ and then the transformed potential 
$V^{\left( \nu \right) }(x;a)$ are singular at the nodes of $\psi _{\nu
}(x;a)$. For instance, if $\psi _{n}(x;a)$ ($\nu =n$) is a bound state of $%
\widehat{H}(a)$, $V^{\left( n\right) }$ is regular only when $n=0$, that is
when $\psi _{n=0}$ is the ground state of $\widehat{H}$, and we recover the
usual SUSY partnership in quantum mechanics. Starting from an excited state,
that is for $n\geq 1$, the transformed potential presents exactly $n$ second
order poles and a priori we cannot use $A\left( w_{n}\right) $ to build a
regular potential. We can however envisage to use any other regular solution
of Eq(\ref{edr4}) as long as it has no zero on the considered real interval $%
I$, even if it does not correspond to a physical state. As shown in the case
of the isotonic oscillator, we can obtain such solutions by using specific
discrete symmetries $\Gamma _{a}$ which are covariance transformations for
the considered family of potentials. $\Gamma _{a}$ acts on the parameters of
the potential and transforms the RS function of a\ physical eigenstate $%
w_{n} $ into a non singular but unphysical RS function $v_{n}(x;a)=\Gamma
_{a}\left( w_{n}(x;a)\right) $ associated to the eigenvalue $\widetilde{E}%
_{n}(a)=\Gamma _{a}\left( E_{n}(a)\right) $. For a solvable TSIP, $w_{n}$
and $v_{n}$ are known in closed form and the regular extended potential (see
Eq(\ref{pottrans}) and Eq(\ref{foDBT}))

\begin{equation}
\widetilde{V}^{\left( n\right) }(x;a)=V(x;a)+2v_{n}^{\prime }(x;a)
\end{equation}%
is then (quasi) isospectral to $V(x;a)$ with eigenstates given by (see Eq(%
\ref{transfoback2}))

\begin{equation}
\left\{ 
\begin{array}{c}
w_{\lambda }^{\left( n\right) }(x;a)=-v_{n}(x;a)+\frac{E_{k}(a)-\widetilde{E}%
_{n}(a)}{v_{n}(x;a)-w_{k}(x;a)} \\ 
\psi _{k}^{\left( n\right) }(x;a)=\exp \left( -\int
dxw_{k}^{(n)}(x;a)\right) \sim \frac{1}{\sqrt{E_{k}\left( a\right) -%
\widetilde{E}_{n}(a)}}\widehat{A}\left( v_{n}\right) \psi _{k}(x;a)%
\end{array}%
\right. ,  \label{foext}
\end{equation}%
for the energy $E_{k}(a)$.

Interestingly, such combinations of Darboux-B\"{a}cklund transformations and
discrete symmetries appears as natural covariance groups for Painlev\'{e}
equations \cite{Adler2}. Very recently another type of discrete symmetries
have been also considered by Plyushchay et al \cite{plyushchay1,plyushchay2}
in a different context.

\subsection{One dimensional harmonic oscillator}

To illustrate this general scheme we consider the well studied example \cite%
{shnol',samsonov,tkachuk,gomez,carinena,fellows,grandati2} of the 1D HO
which is simplest exceptional first category TSIP. The corresponding
potential with zero ground level$\ $($E_{0}(\omega )=0$) is given by

\begin{equation}
V(x,\omega )=\frac{\omega ^{2}}{4}x^{2}-\frac{\omega }{2}.
\end{equation}

Its spectrum is well known

\begin{equation}
E_{n}\left( \omega \right) =n\omega ;\quad \psi _{n}\left( x\right) \sim
H_{n}\left( \sqrt{\omega /2}x\right) \exp \left( -\omega x^{2}/4\right)
\end{equation}%
and the corresponding RS functions $w_{n}(x)$ can be written as terminating
continued fractions \cite{grandati} as%
\begin{equation}
w_{n}(x,\omega )=w_{0}(x,\omega )+R_{n}(x,\omega ),  \label{RS functions OH1}
\end{equation}%
where%
\begin{equation}
\left\{ 
\begin{array}{c}
w_{0}(x,\omega )=\frac{\omega }{2}x \\ 
R_{n}(x,\omega )=-\frac{n\omega }{\omega x-}\Rsh ...\Rsh \frac{\left(
n-j+1\right) \omega }{\omega x-}\Rsh ...\Rsh \frac{1}{x}=-\left( \log
H_{n}\left( \sqrt{\omega /2}x\right) \right) ^{\prime }.%
\end{array}%
\right.  \label{RS functions OH2}
\end{equation}

The unique parameter transformation which preserves the functional form $%
V(x,\omega )$ is the $\omega $ inversion%
\begin{equation}
\omega \overset{\Gamma _{\omega }}{\rightarrow }\left( -\omega \right)
,\left\{ 
\begin{array}{c}
V(x;\omega )\overset{\Gamma _{\omega }}{\rightarrow }V(x;\omega )+\omega \\ 
w_{n}(x;\omega )\overset{\Gamma _{\omega }}{\rightarrow }v_{n}(x;\omega
)=w_{n}(x;-\omega ),%
\end{array}%
\right.
\end{equation}%
$v_{n}(x;\omega )$ satisfying

\begin{equation}
-v_{n}^{\prime }(x;\omega )+v_{n}^{2}(x;\omega )=V(x;\omega )-E_{-\left(
n+1\right) }\left( \omega \right) ,  \label{oregpot}
\end{equation}%
that is, $E_{n}\left( \omega \right) \overset{\Gamma _{\omega }}{\rightarrow 
}E_{-\left( n+1\right) }\left( \omega \right) $. From Eq.(\ref{RS functions
OH1}) and Eq.(\ref{RS functions OH2}) we deduce 
\begin{equation}
v_{n}(x;\omega )=v_{0}(x;\omega )+Q_{n}(x;\omega ),  \label{oregRSfct}
\end{equation}%
with

\begin{equation}
v_{0}(x;\omega )=-\frac{\omega }{2}x  \label{oregRSfct2}
\end{equation}%
and

\begin{eqnarray}
Q_{n}(x;\omega ) &=&-\frac{n\omega }{\omega x+}\Rsh ...\Rsh \frac{\left(
n-j+1\right) \omega }{\omega x+}\Rsh ...\Rsh \frac{1}{x}  \label{oregRSfct3}
\\
&=&-\left( \log H_{n}\left( i\sqrt{\omega /2}x\right) \right) ^{\prime }. 
\notag
\end{eqnarray}

Clearly, $Q_{n}(x;\omega )$ does not present any singularity on the real
line, except possibly one at the origin. Indeed the terminating continued
fraction has only positive terms which implies that there is no positive
singularity and then, since the potential has a even parity, any singularity
on the whole real axis, except one at the origin when the number $n$ of
denominators is odd. This can be recovered more directly from the expression
of $Q_{n}$ in terms of Hermite polynomials of imaginary argument since the
Hermite polynomials have all their zeros on the real line, with a zero at
the origin for odd $n$. Using the correspondence between Hermite and
Laguerre polynomials given by

\begin{equation}
H_{n}\left( i\sqrt{\omega /2}x\right) =\left\{ 
\begin{array}{c}
\left( -1\right) ^{m}2^{2m}m!L_{m}^{-1/2}\left( -\omega x^{2}/2\right)
,\quad n=2m \\ 
\left( -1\right) ^{m}2^{2m+1}m!i\sqrt{\omega /2}xL_{m}^{1/2}\left( -\omega
x^{2}/2\right) ,\quad n=2m+1,%
\end{array}%
\right.  \label{Herm_Lag}
\end{equation}%
the regularity properties of are direct consequences of the
Kienast-Lawton-Hahn theorem \cite{szego,magnus} which establishes

\emph{Kienast-Lawton-Hahn's Theorem }

\ \ \ \ \ Suppose that $\alpha \notin \left\{ -n,...,-1\right\} $. Then $%
\mathit{L}_{n}^{\left( \alpha \right) }\left( z\right) $ admits

\ \ \ \ \ \ \ \ \ \ \ \ 1) $n$ positive zeros if $\alpha >-1$

\ \ \ \ \ \ \ \ \ \ \ \ 2) $n+\left[ \alpha \right] +1$ positive zeros if $%
-n<\alpha <-1$ ($\left[ \left\vert \alpha \right\vert \right] $ means the
integer part of $\alpha $)

\ \ \ \ \ \ \ \ \ \ \ \ 3) No positive zero if $\alpha <-n$

\ \ \ \ The number of negative zeros is always $0$ or $1$.

\ \ \ \ \ \ \ \ \ \ \ \ 1) $0$ if $\alpha >-1$

\ \ \ \ \ \ \ \ \ \ \ \ 2) $0$ if $-2k-1<\alpha <-2k$ and $1$ if $-2k<\alpha
<-2k+1$, with $-n<\alpha <-1$

\ \ \ \ \ \ \ \ \ \ \ \ 3) $0$ if $n$ is even and $1$ if $n$ is odd, with $%
\alpha <-n$

\ \ \ \ Only when $\alpha \in \left\{ -n,...,-1\right\} ,$ we have a zero of 
$\mathit{L}_{n}^{\left( \alpha \right) }\left( z\right) $ at the origin with
multiplicity $\left\vert \alpha \right\vert $. If $\alpha $ decreases
through an odd value in $\left\{ -n,...,-1\right\} $, a negative zero is
gained and a positive one is lost. If the crossed value is even,
simultaneously two zeros,\ one negative and one positive, disappear. \ \ \ \
\ \ \ \ \ \ \ \ 

Applying the DBT $A\left( v_{n}\right) $ to $w_{k}$ (see Eq(\ref%
{transfoback2})), we obtain

\begin{equation}
w_{k}(x;\omega )\overset{A\left( v_{n}\right) }{\rightarrow }w_{k}^{\left(
n\right) }(x;\omega )=-v_{n}(x;\omega )+\frac{E_{n+1+k}(\omega )}{%
v_{n}(x;\omega )-w_{k}(x;\omega )},  \label{backOH}
\end{equation}%
where $w_{k}^{\left( n\right) }(x;\omega )$ is an RS function at energy $%
E_{k}(\omega )$ for the extended potential

\begin{equation}
V^{\left( n\right) }(x;\omega )=V(x;\omega )+2v_{n}^{\prime }(x;\omega
)=V(x;\omega )-\omega +2Q_{n}^{\prime }(x;\omega ).  \label{oregSUSYpart}
\end{equation}

We recover here the results obtained in \cite%
{shnol',samsonov,tkachuk,gomez,carinena,fellows,grandati2}. In particular,
for $n=1$ $\widetilde{V}^{\left( 1\right) }(x)$ is the $l=1$ isotonic
potential

\begin{equation}
V^{\left( 1\right) }(x;\omega )=V(x;\omega )-\omega +\frac{2}{x^{2}}=\frac{%
\omega ^{2}}{4}x^{2}+\frac{2}{x^{2}}-\frac{3\omega }{2}
\end{equation}%
and for $n=2$, $V^{\left( 2\right) }(x;\omega )$ is the CPRS \cite{carinena}
potential

\begin{equation}
V^{\left( 2\right) }(x;\omega )=\frac{\omega ^{2}}{4}x^{2}+4\omega \frac{%
\omega ^{2}x^{2}-1}{\left( \omega ^{2}x^{2}+1\right) ^{2}}-\frac{3}{2}\omega
.
\end{equation}

For every $n\geq 0$, $V^{\left( n\right) }(x;\omega )$ is (quasi)isospectral
to $V(x;\omega )$

\begin{equation}
V^{\left( n\right) }(x;\omega )\underset{iso}{\equiv }V(x;\omega ).
\label{oregSUSYpart2}
\end{equation}%
and regular on the real line $\mathbb{R}$ if $n$ is even and on the positive
half real line $\mathbb{R}^{+\ast }$ if $n$ is odd. To keep the same
definition domain for the initial and extended potentials, we then must
consider only even values of $n=2m$.

The isospectrality established above is not strict. Indeed, we have clearly

\begin{equation}
v_{n}^{\prime }(x;\omega )+v_{n}^{2}(x;\omega )=V^{\left( n\right)
}(x;\omega )-E_{-\left( n+1\right) }\left( \omega \right) ,
\end{equation}%
that is, $-v_{n}(x;\omega )$ is a regular RS function for the extended
potential $V^{\left( n\right) }(x;\omega )$, associated to the eigenvalue $%
E_{-\left( n+1\right) }\left( \omega \right) <0$. Then

\begin{equation}
\psi _{-}^{\left( 2m\right) }(x;\omega )\sim \exp \left( +\int
v_{2m}(x;\omega )dx\right) =\frac{\exp \left( -\frac{\omega x^{2}}{4}\right) 
}{H_{2m}\left( i\sqrt{\omega /2}x\right) }\sim \frac{\exp \left( -\frac{%
\omega x^{2}}{4}\right) }{L_{m}^{-1/2}\left( -\omega x^{2}/2\right) }
\end{equation}%
is a physical eigenstate of $\widehat{H}^{\left( 2m\right) }$ and more
precisely its fundamental state. Consequently the superpartner of the
extended potential $V^{\left( 2m\right) }(x;\omega )$ is

\begin{equation}
\widetilde{V}^{\left( 2m\right) }(x;\omega )=V^{\left( 2m\right) }(x;\omega
)-2v_{2m}^{\prime }(x;\omega )=V(x;\omega ),\ m\geq 1  \label{SUSYpartHerm}
\end{equation}%
and we recover the fact that the DBT $A\left( v_{n}\right) $ is a backward
SUSY partnership.

The eigenfunctions of $\widehat{H}^{\left( 2m\right) }(\omega
)=-d^{2}/dx^{2}+V^{\left( 2m\right) }(x;\omega )$ corresponding to the
energies $E_{k}(\omega )$ are given by

\begin{equation}
\psi _{k}^{\left( 2m\right) }(x;\omega )=\exp \left( -\int
dxw_{k}^{(2m)}(x;a)\right) \sim \frac{1}{\sqrt{E_{2m+1+k}(\omega )}}\widehat{%
A}\left( v_{2m}\right) \psi _{k}(x;\omega ),
\end{equation}%
that is, using Eq(\ref{Herm_Lag})

\begin{equation}
\psi _{k}^{\left( 2m\right) }(x;\omega )\sim P_{\left( m,k\right) }(x)\frac{%
\exp \left( -\omega x^{2}/4\right) }{L_{m}^{-1/2}\left( -\omega
x^{2}/2\right) },  \label{foextHO}
\end{equation}%
where the polynomials

\begin{equation}
P_{\left( m,k\right) }(x)=\frac{1}{2}L_{m}^{-1/2}\left( -\omega
x^{2}/2\right) H_{k+1}\left( \sqrt{\omega /2}x\right) +\sqrt{\omega /2}%
xL_{m-1}^{1/2}\left( -\omega x^{2}/2\right) H_{k}\left( \sqrt{\omega /2}%
x\right)  \label{polHerm}
\end{equation}%
of respective degrees $2m+k+1$ constitute, with the constant $1$, an
orthogonal family on the real line with respect to the weight

\begin{equation}
w_{m}(x)=\frac{\exp \left( -\omega x^{2}/2\right) }{\left(
L_{m}^{-1/2}\left( -\omega x^{2}/2\right) \right) ^{2}}.  \label{poidsHerm}
\end{equation}

Note that for $k=2l,\ l\in \mathbb{N}$, we have

\begin{eqnarray}
P_{\left( m,2l\right) }(x) &\sim &\sqrt{\omega /2}x\left( L_{m}^{-1/2}\left(
-\omega x^{2}/2\right) L_{l}^{1/2}\left( \omega x^{2}/2\right)
+L_{m-1}^{1/2}\left( -\omega x^{2}/2\right) L_{l}^{-1/2}\left( \omega
x^{2}/2\right) \right) \\
&=&\sqrt{\omega /2}xL_{l}^{1/2}\left( \omega x^{2}/2\right)  \notag
\end{eqnarray}%
where $L_{l}^{1/2}\left( z\right) $ is an EOP of the $L1$ series \cite%
{gomez4}. This is coherent with the fact that the 1D HO on is obtained as
the singular limit at $a\rightarrow 0$ of the isotonic potential

\begin{equation}
V(x;\omega ,a)=\frac{\omega ^{2}}{4}x^{2}+\frac{a(a-1)}{x^{2}}-\omega \left(
a+\frac{1}{2}\right) .  \label{isotpot}
\end{equation}

For $a>0$ the presence of the centrifugal barrier restricts the definition
domain to the positive half line$\ x>0$ and the energy spectrum include only
the level $E_{k}\left( \omega \right) =k\omega $ associated to an even
quantum number $k=2l$. The $L1$ series of rational extensions of $V(x;\omega
,a)$ is then built using DBT based on RS functions actually regularized via
the same $\omega $ inversion $\Gamma _{\omega }$ that we used above for the
HO \cite{grandati3}.

Finally, for the odd values of the quantum number $k=2l+1$ we can write

\begin{equation}
P_{\left( m,2l+1\right) }(x)\sim \left( l+1\right) L_{m}^{-1/2}\left(
-\omega x^{2}/2\right) L_{l+1}^{-1/2}\left( \omega x^{2}/2\right) -\sqrt{%
\omega /2}xL_{m-1}^{1/2}\left( -\omega x^{2}/2\right) L_{l}^{1/2}\left(
\omega x^{2}/2\right) .
\end{equation}

\section{Morse potential}

The Morse potential with zero ground level ($E_{0}(a,b)=0$) is the second
exceptional primary TSIP of the first category \cite{grandati}. It is given
by \cite{morse,cooper,Dutt}

\begin{equation}
V(y;a,b)=b^{2}y^{2}-2\left( a+\frac{\alpha }{2}\right) by+a^{2},\ a,b>0
\label{potMorse}
\end{equation}%
where $y=\exp \left( -\alpha x\right) >0,\ x\in \mathbb{R}$. It possesses
exactly $\left[ a\right] $ bound states ( $\left[ a\right] $ being the
integer part of $a$) which are given by

\begin{equation}
\psi _{n}\left( x;a,b\right) \sim y^{a/\alpha -n}e^{-by/\alpha }\mathit{L}%
_{n}^{2(a/\alpha -n)}(2by/\alpha ),\ n\in \left\{ 0,...,\left[ a\right]
-1\right\} ,  \label{foMorse}
\end{equation}%
with the corresponding energies $E_{n}(a)=a^{2}-a_{n}^{2},$ where $%
a_{k}=a-k\alpha $.

In terms of the $y$ variable, the associated RS equation is

\begin{equation}
\alpha yw_{n}^{\prime }(y;a,b)+w_{n}^{2}(y;a,b)=V(y;a,b)-E_{n}\left( a\right)
\end{equation}%
\qquad and its solutions associated to the physical eigenstates Eq(\ref%
{foMorse}) are

\begin{equation}
w_{n}(y;a,b)=w_{0}(y;a,b)+R_{n}(y;a,b),  \label{RS functions Morse1}
\end{equation}%
where%
\begin{equation}
w_{0}(y;a,b)=-by+a  \label{RS functions Morse2}
\end{equation}%
and

\begin{eqnarray}
R_{n}(y;a,b) &=&-\frac{E_{n}\left( a\right) }{a+a_{1}-2by-}\Rsh ...\Rsh 
\frac{E_{n}\left( a\right) -E_{j-1}\left( a\right) }{a_{j-1}+a_{j}-2by-}\Rsh
...\Rsh \frac{E_{n}\left( a\right) -E_{n-1}\left( a\right) }{%
a_{n-1}+a_{n}-2by}  \label{RS functions Morse3} \\
&=&-n\alpha +\alpha y\left( \log \mathit{L}_{n}^{2(a/\alpha -n)}(2by/\alpha
)\right) ^{\prime }.  \notag
\end{eqnarray}

The only parameters transformation under which the Morse potential Eq(\ref%
{potMorse}) is covariant, is

\begin{equation}
\left( a,b\right) \overset{\Gamma _{a,b}}{\rightarrow }\left( \underset{%
-a_{-1}}{\underbrace{-a-1}},-b\right) ,\left\{ 
\begin{array}{c}
V(x;a,b)\overset{\Gamma _{a,b}}{\rightarrow }V(x;a,b)-E_{-1}\left( a\right)
\\ 
w_{n}(x;a,b)\overset{\Gamma _{a,b}}{\rightarrow }%
v_{n}(x;a,b)=w_{n}(x;-a_{-1},-b),%
\end{array}%
\right.  \label{discreteMorse}
\end{equation}%
where

\begin{equation}
\alpha yv_{n}^{\prime }(y;a,b)+v_{n}^{2}(y;a,b)=V(y;a,b)-E_{-(n+1)}\left(
a\right) ,
\end{equation}%
since $a_{k}\overset{\Gamma _{a,b}}{\rightarrow }a_{-(k+1)}$ and $%
E_{n}\left( a\right) \overset{\Gamma _{a,b}}{\rightarrow }%
a_{-1}^{2}-a_{-(n+1)}^{2}=E_{-(n+1)}\left( a\right) -E_{-1}\left( a\right) $.

From Eq.(\ref{RS functions Morse2}) and Eq.(\ref{RS functions Morse3}), we
deduce

\begin{equation}
v_{n}(x;a,b)=v_{0}(x;a,b)+Q_{n}(x;a,b),  \label{MregRSfct1}
\end{equation}%
where

\begin{equation}
v_{0}(y,a,b)=by-a_{-1}  \label{MregRSfct2}
\end{equation}%
and

\begin{eqnarray}
Q_{n}(y,a,b) &=&-\frac{E_{-(n+1)}\left( a\right) -E_{-1}\left( a\right) }{%
-\left( a_{-1}+a_{-2}\right) +2by-}\Rsh ...\Rsh \frac{E_{-(n+1)}\left(
a\right) -E_{-j}\left( a\right) }{-\left( a_{-j}+a_{-j-1}\right) +2by-}\Rsh
...\Rsh \frac{E_{-(n+1)}\left( a\right) -E_{-n}\left( a\right) }{-\left(
a_{-n}+a_{-n-1}\right) +2by}  \label{MregRSfct3} \\
&=&-n\alpha +\alpha y\left( \log \mathit{L}_{n}^{-2(a/\alpha
+1+n)}(-2by/\alpha )\right) ^{\prime }.  \notag
\end{eqnarray}

The Kienast-Lawton-Hahn's theorem ensures that for even values of $n$, $%
Q_{n}(y,a,b)$ and then $v_{n}(x;a,b)$ are regular for every $y>0$, that is,
every $x\in \mathbb{R}$. Applying the DBT $A\left( v_{n}\right) $ (see Eq(%
\ref{transfoback2})) to $w_{k}$ gives

\begin{equation}
w_{k}(x;a,b)\overset{A\left( v_{n}\right) }{\rightarrow }w_{k}^{\left(
n\right) }(x;a,b)=-v_{n}(x;a,b)+\frac{E_{k}\left( a\right) -E_{-(n+1)}\left(
a\right) }{v_{n}(x;a,b)-w_{k}(x;a,b)},
\end{equation}%
where $w_{k}^{\left( n\right) }(x;\omega )$ satisfies

\begin{equation}
-w_{k}^{\left( n\right) \prime }(x;a,b)+\left( w_{k}^{\left( n\right)
}(x;a,b)\right) ^{2}=V^{\left( n\right) }(x;a,b)-E_{k}\left( a\right) ,
\end{equation}%
with

\begin{equation}
V^{\left( n\right) }(x;a,b)=V(x;a,b)+2v_{n}^{\prime
}(x;a,b)=V(y;a_{-1},b)+E_{-1}\left( a\right) -2\alpha yQ_{n}^{\prime
}(y;a,b).  \label{regextMorse}
\end{equation}%
\ 

In the following we consider the case where $n$ takes even values $n=2m$. $%
V^{\left( 2m\right) }(x;a,b)$ is then regular on the positive half line and
isospectral to $V(x;a,b)$

\begin{equation}
V^{\left( 2m\right) }(x;a,b)\underset{iso}{\equiv }V(x;a,b)
\end{equation}

Again, as in the preceding case, the isospectrality is not strict since \ \
\ \ \ \ \ 

\begin{equation}
v_{n}^{\prime }(x;a,b)+v_{n}^{2}(x;a,b)=V^{\left( 2m\right)
}(x;a,b)-E_{-(n+1)}\left( a\right) ,
\end{equation}%
that is, $-v_{n}(x;a,b)$ is a regular RS function for the extended potential 
$V^{\left( 2m\right) }(x;a,b)$, associated to the eigenvalue $%
E_{-(n+1)}\left( a\right) <0$. The asymptotic behaviour of the corresponding
eigenstate is

\begin{equation}
\psi _{-}^{\left( 2m\right) }(x;a,b)=\exp \left( +\int
v_{2m}(x;a,b)dx\right) \underset{x\rightarrow \pm \infty }{\sim }e^{-\left(
a+\left( 2m+1\right) \alpha \right) x}\exp \left( -\frac{b}{\alpha }%
e^{-\alpha x}\right)  \label{fondextMorse}
\end{equation}%
from which we deduce that $\psi _{-}^{\left( 2m\right) }$ is the fundamental
state for $H^{\left( 2m\right) }$. The superpartner of the extended
potential $V^{\left( 2m\right) }(x;a,b)=V(x;a,b)+2v_{2m}^{\prime }(x;a,b)$
is then defined as

\begin{equation}
\widetilde{V}^{\left( 2m\right) }(x;a,b)=V^{\left( 2m\right)
}(x;a,b)+2\left( -v_{2m}^{\prime }(x;a,b)\right) =V(x;a,b),\ n\geq 1
\end{equation}%
and the DBT $A\left( v_{2m}\right) $ is a backward SUSY partnership. We
recover here the results obtained by G\'{o}mez-Ullate, Kamran and Milson 
\cite{gomez} in a different way.

The excited physical eigenstate\ of $\widehat{H}^{\left( 2m\right)
}(a,b)=-d^{2}/dx^{2}+V^{\left( 2m\right) }(x;a,b)$ at the energy $%
E_{k}\left( a\right) ,\ k\geq 0,$ is given by (see Eq(\ref{foext}))

\begin{equation}
\psi _{k}^{\left( 2m\right) }(x;a,b)=\exp \left( \frac{1}{\alpha }\int dy%
\frac{w_{k}^{\left( 2m\right) }(y;a,b)}{y}\right) \sim \frac{1}{\sqrt{%
E_{k}\left( a\right) -E_{-(2m+1)}\left( a\right) }}\widehat{A}\left(
v_{2m}\right) \psi _{k}(x;a,b).  \label{foextMorse}
\end{equation}

Inserting Eq(\ref{MregRSfct2}), Eq(\ref{MregRSfct3}) and Eq(\ref{foMorse})
into Eq(\ref{foextMorse}) and using the following identities for GLP

\begin{equation}
\left\{ 
\begin{array}{c}
\mathit{L}_{n}^{\left( \beta \right) }\left( z\right) +\mathit{L}%
_{n-1}^{\left( \beta +1\right) }\left( z\right) =\mathit{L}_{n}^{\left(
\beta +1\right) }\left( z\right) \\ 
z\mathit{L}_{n-1}^{\left( \beta +1\right) }\left( z\right) =(n+\beta )%
\mathit{L}_{n-1}^{\left( \beta \right) }\left( z\right) -n\mathit{L}%
_{n}^{\left( \beta \right) }\left( z\right) ,%
\end{array}%
\right.
\end{equation}%
we obtain (in order to simplify the expressions we fix the $x$ scale such
that $\alpha =1$)

\begin{equation}
\psi _{k}^{\left( 2m\right) }(x;a,b)\sim M_{a,k}^{(2m)}\left( z\right) \frac{%
z^{a-k}e^{-z/2}}{\mathit{L}_{2m}^{-2(a+1+2m)}(-z)},\ \psi _{-}^{\left(
2m\right) }(x;a,b)\sim \frac{z^{a+1+2m}e^{-z/2}}{\mathit{L}%
_{2m}^{-2(a+1+2m)}(-z)}  \label{foextMorse2}
\end{equation}%
where $z=2by$ and

\begin{equation}
M_{a,k}^{(2m)}\left( z\right) =2\left( m+a+1\right) \mathit{L}%
_{2m-1}^{-2(a+2m+1)}(-z)\mathit{L}_{k}^{2(a-k)}(z)-\left( k+1\right) \mathit{%
L}_{k+1}^{2(a-k)}(z)\mathit{L}_{2m}^{-2(a+2m+1)}(-z)  \label{polyM1}
\end{equation}%
which is a polynomial of degree $2m+k+1$ with

\begin{equation}
M_{a,k}^{(2m)}\left( 0\right) =-\frac{\left( 2a+2m+2\right) _{2m}\left(
2a-2k+1\right) _{k}}{(2m)!k!},
\end{equation}%
$\left( a\right) _{n}$ being the usual Pochhammer function $\left( a\right)
_{n}=a(a+1)...(a+n-1)$ \cite{magnus}$.$

From the orthonormality conditions $<\psi _{k}^{\left( 2m\right)
}(x;a,b)\mid \psi _{k^{\prime }}^{\left( 2m\right) }(x;a,b)>=\delta
_{k,k^{\prime }}$ we deduce that the polynomials

\begin{equation}
\left\{ 
\begin{array}{c}
B_{-}^{\left( 2m\right) }\left( z,a\right) =1 \\ 
B_{k}^{\left( 2m\right) }\left( z,a\right) =z^{k+2m+1}M_{a,k}^{(2m)}\left( 
\frac{1}{z}\right) ,\ k\in \left\{ 0,...,\left[ a\right] -1\right\} ,%
\end{array}%
\right.
\end{equation}%
are orthogonal on the positive half line with respect to the weight

\begin{equation}
w^{\left( 2m\right) }\left( z,a\right) =\frac{e^{-1/z}}{z^{2\left(
a+2m\right) +3}\left( \mathit{L}_{2m}^{-2(a+1+2m)}(-1/z)\right) ^{2}}.
\end{equation}

\section{Radial effective Kepler-Coulomb}

The effective radial Kepler-Coulomb (ERKC) potential with zero ground level (%
$E_{0}(a)=0$) is the third and last exceptional primary TSIP of the first
category \cite{grandati}. It is defined on the positive half line as%
\begin{equation}
V(x;a)=\frac{a(a-1)}{x^{2}}-\frac{\gamma }{x}+V_{0}\left( a\right) ,\ \gamma
>0,\ a>1  \label{potKC}
\end{equation}%
where $x>0$ and $V_{0}\left( a\right) =\gamma ^{2}/4a^{2}$.

Its bound states are given by

\begin{equation}
\psi _{n}\left( x;a\right) =\exp \left( -\int dxw_{n}(x;a)\right) \sim
x^{a}e^{-\gamma x/2a_{n}}\mathit{L}_{n}^{\left( 2a-1\right) }(\gamma
x/a_{n}),\ n\geq 0,  \label{foKC}
\end{equation}%
with the corresponding energies $E_{n}(a)=V_{0}\left( a\right) -V_{0}\left(
a_{n}\right) $, where $a_{k}=a+k.$

The associated RS equation is%
\begin{equation}
-w_{n}^{\prime }(x;a)+w_{n}^{2}(x;a)=V(x;a)-E_{n}\left( a\right)
\label{RSeqMorse}
\end{equation}

The solutions of eq(\ref{RSeqMorse}) corresponding to the physical
eigenstates are given by

\begin{equation}
w_{n}(x;a)=w_{0}(x;a)+R_{n}(x;a),  \label{RS functions KC1}
\end{equation}%
where

\begin{equation}
w_{0}(x;a)=-\frac{a}{x}+\gamma /2a  \label{RS functions KC2}
\end{equation}%
and

\begin{eqnarray}
R_{n}(y;a) &=&-\frac{E_{n}\left( a\right) }{w_{0}(x;a)+w_{0}(y;a_{1})-}\Rsh
...\Rsh \frac{E_{n}\left( a\right) -E_{j-1}\left( a\right) }{%
w_{0}(x;a_{j-1})+w_{0}(x;a_{j})-}\Rsh ...\Rsh \frac{E_{n}\left( a\right)
-E_{n-1}\left( a\right) }{w_{0}(x;a_{n-1})+w_{0}(x;a_{n})}
\label{RS functions KC3} \\
&=&\frac{\gamma }{2a_{n}}-\frac{\gamma }{2a}-\left( \log \left( \mathit{L}%
_{n}^{\left( 2a-1\right) }(\gamma x/a_{n})\right) \right) ^{\prime }.  \notag
\end{eqnarray}

The only covariance transformation for the ERKC potentials is given by

\begin{equation}
a\overset{\Gamma _{a}}{\rightarrow }\underset{-a_{-1}}{\underbrace{1-a}}%
,\left\{ 
\begin{array}{c}
V(x;a)\overset{\Gamma _{a}}{\rightarrow }V(x;a)-E_{-1}\left( a\right) \\ 
w_{n}(x;a)\overset{\Gamma _{a}}{\rightarrow }v_{n}(x;a)=w_{n}(x;-a_{-1}),%
\end{array}%
\right.  \label{discreteKC}
\end{equation}%
with 
\begin{equation}
a_{k}\overset{\Gamma _{a}}{\rightarrow }1-a+k=-a_{-\left( k+1\right) },\quad
E_{n}\left( a\right) \overset{\Gamma _{a}}{\rightarrow }\gamma ^{2}/4\left( 
\frac{1}{a_{-1}^{2}}-\frac{1}{a_{-(n+1)}^{2}}\right) =E_{-(n+1)}\left(
a\right) -E_{-1}\left( a\right) .
\end{equation}

We then have

\begin{equation}
-v_{n}^{\prime }(x;a)+v_{n}^{2}(x;a)=V(x;a)-E_{-(n+1)}\left( a\right)
\end{equation}%
and from Eq.(\ref{RS functions KC2}) and Eq.(\ref{RS functions KC3}), we
deduce

\begin{equation}
v_{n}(x;a)=v_{0}(x;a)+Q_{n}(x;a),  \label{KCregRSfct1}
\end{equation}%
where

\begin{equation}
\left\{ 
\begin{array}{c}
v_{0}(x,a)=\frac{a_{-1}}{x}-\frac{\gamma }{2a_{-1}} \\ 
Q_{n}(x,a)=-\frac{\gamma }{2a_{-(n+1)}}+\frac{\gamma }{2a_{-1}}-\left( \log
\left( \mathit{L}_{n}^{\left( 1-2a\right) }(-\gamma x/a_{-(n+1)})\right)
\right) ^{\prime }.%
\end{array}%
\right.  \label{KCregRSfct2}
\end{equation}

If the argument of the GLP is positive, that is, if $a<n+1$ the
Kienast-Lawton-Hahn theorem ensures that $Q_{n}(x,a)$ is regular for $x>0$
if $1-2a<-n$, that is, if

\begin{equation}
\frac{n+1}{2}<a<n+1
\end{equation}

Another possibility to ensure the regularity of $Q_{n}(x,a)$ is to consider
values of $a$ such that $a>n+1,$ where the argument of the GLP is now
negative. From the Kienast-Lawton-Hahn theorem, we then deduce that in this
case for each even value of $n=2m$, $Q_{2m}(x,a)$ is regular.

The DBT $A\left( v_{n}\right) $ applied to $w_{k}$ gives

\begin{equation}
w_{k}(x;a)\overset{A\left( v_{n}\right) }{\rightarrow }w_{k}^{\left(
n\right) }(x;a)=-v_{n}(x;a)+\frac{E_{k}\left( a\right) -E_{-(n+1)}\left(
a\right) }{v_{n}(x;a)-w_{k}(x;a)},  \label{DBTKC}
\end{equation}%
where $w_{k}^{\left( n\right) }(x;a)$ satisfies

\begin{equation}
-w_{k}^{\left( n\right) \prime }(x;a)+\left( w_{k}^{\left( n\right)
}(x;a)\right) ^{2}=V^{\left( n\right) }(x;a)-E_{k}\left( a\right) ,
\end{equation}%
with

\begin{equation}
V^{\left( n\right) }(x;a)=V(x;a)+2v_{n}^{\prime
}(x;a)=V(x;a_{-1})+E_{-1}\left( a\right) +2Q_{n}^{\prime }(x;a).
\end{equation}

In the cases where

\begin{equation}
\left\{ 
\begin{array}{c}
\frac{n+1}{2}<a<n+1\ \qquad \quad \quad \left( i\right) \\ 
n=2m,\quad a>n+1,\text{ \ \ \ \ \ }\left( ii\right)%
\end{array}%
\right.  \label{condregKC}
\end{equation}%
\ $V^{\left( n\right) }(x;a)$ is regular on the positive half line and
isospectral to $V(x;a)$

\begin{equation}
V^{\left( n\right) }(x;a)\underset{iso}{\equiv }V(x;a).
\end{equation}

We have also

\begin{equation}
v_{n}^{\prime }(x;a)+v_{n}^{2}(x;a)=V^{\left( n\right)
}(x;a)-E_{-(n+1)}\left( a\right) ,
\end{equation}%
that is, $-v_{n}(x;a)$ is a regular RS function for the extended potential $%
V^{\left( n\right) }(x;a)$, associated to the eigenvalue $E_{-(n+1)}\left(
a\right) <0$, when $\frac{n+1}{2}<a$. Moreover

\begin{equation}
\psi _{-}^{\left( n\right) }(x;a)=\exp \left( +\int v_{n}(x;a)dx\right) \sim 
\frac{x^{a-1}\exp \left( -\frac{\gamma }{2a_{-\left( n+1\right) }}x\right) }{%
\mathit{L}_{n}^{\left( 1-2a\right) }(-\gamma x/a_{-\left( n+1\right) })}.
\label{fondextKC}
\end{equation}

In the case $(ii)$ (see Eq.(\ref{condregKC})), $a_{-\left( 2m+1\right) }>0$
and $\psi _{-}^{\left( 2m\right) }$ is a physical eigenstate for $\widehat{H}%
^{\left( 2m\right) }$ with the lowest eigenvalue. In other words, $\psi
_{-}^{\left( 2m\right) }$ is the fundamental state for $\widehat{H}^{\left(
2m\right) }$ and, as for the two preceding exceptional primary TSIP of the
first category, the isospectrality is not strict. On the other hand, in the
case $(i)$ ($a_{-\left( n+1\right) }<0$), $\psi _{-}^{\left( n\right) }$ is
not in the physical spectrum and in this regime the isospectrality between $%
\widehat{H}^{\left( n\right) }$ and $\widehat{H}$ becomes strict.

Consider first the case $\left( ii\right) $. The superpartner of the
extended potential $V^{\left( 2m\right) }(x;a)=V(x;a)+2v_{2m}^{\prime }(x;a)$
is given by

\begin{equation}
\widetilde{V}^{\left( 2m\right) }(x;a)=V^{\left( 2m\right) }(x;a)+2\left(
-v_{2m}^{\prime }(x;a)\right) =V(x;a)
\end{equation}%
and the DBT $A\left( v_{2m}\right) $ corresponds to a backward SUSY
partnership.

The fundamental eigenstate of $\widehat{H}^{\left( 2m\right)
}(a)=-d^{2}/dx^{2}+V^{\left( 2m\right) }(x;a)$ at the energy $%
E_{-(2m+1)}\left( a\right) $ is

\begin{equation}
\psi _{-}^{\left( 2m\right) }(x;a)\sim \frac{x^{a-1}\exp \left( -\frac{%
\gamma x}{2\left\vert a_{-\left( 2m+1\right) }\right\vert }\right) }{\mathit{%
L}_{2m}^{\left( 1-2a\right) }(-\gamma x/\left\vert a_{-(2m+1)}\right\vert )}
\end{equation}%
and the excited eigenstates at energy $E_{k}\left( a\right) ,\ k\geq 0$ are
(see Eq.(\ref{foDBT}))

\begin{equation}
\psi _{k}^{\left( 2m\right) }(x;a)\sim \frac{1}{\sqrt{E_{k}\left( a\right)
-E_{-(2m+1)}\left( a\right) }}\widehat{A}\left( v_{2m}\right) \psi _{k}(x;a),
\label{foextKC}
\end{equation}%
that is,

\begin{equation}
\psi _{k}^{\left( 2m\right) }(x;a)\sim x^{a-1}e^{-\gamma x/2a_{k}}\frac{%
N_{a,k}^{(2m)}\left( x\right) }{\mathit{L}_{2m}^{\left( 1-2a\right)
}(-\gamma x/\left\vert a_{-(2m+1)}\right\vert )},  \label{foextKC1}
\end{equation}%
where

\begin{eqnarray}
N_{a,k}^{(n)}\left( x\right) &=&\left( 1-2a\right) \mathit{L}_{k}^{\left(
2a-1\right) }(\gamma x/a_{k})\mathit{L}_{n}^{\left( 1-2a\right) }(-\gamma
x/a_{-(n+1)}) \\
&&+\left( a-\frac{n+1}{2}\right) \mathit{L}_{k}^{\left( 2a-1\right) }(\gamma
x/a_{k})\mathit{L}_{n}^{\left( -2a\right) }(-\gamma x/a_{-(n+1)})+\left( a+%
\frac{k-1}{2}\right) \mathit{L}_{k}^{\left( 2a-2\right) }(\gamma x/a_{k})%
\mathit{L}_{n}^{\left( 1-2a\right) }(-\gamma x/a_{-(n+1)})  \notag \\
&&+\frac{k+1}{2}\mathit{L}_{k+1}^{\left( 2a-1\right) }(\gamma x/a_{k})%
\mathit{L}_{n}^{\left( 1-2a\right) }(-\gamma x/a_{-(n+1)})-\frac{n+1}{2}%
\mathit{L}_{k}^{\left( 2a-1\right) }(\gamma x/a_{k})\mathit{L}_{n+1}^{\left(
-2a\right) }(-\gamma x/a_{-(n+1)}),  \notag
\end{eqnarray}%
is a polynomial of degree $n+k+1$. From the orthonormality condition of the
eigenstates of $\widehat{H}^{\left( 2m\right) }(a)$ we obtain that the
function\bigskip s $C_{-}^{\left( 2m\right) }\left( x,a\right) =1$ and%
\begin{equation}
C_{k}^{\left( 2m\right) }\left( x,a\right) =e^{-\gamma
x/2a_{k}}N_{a,k}^{(2m)}\left( x\right) ,\ k\geq 0,
\end{equation}%
constitute an orthogonal family on the positive half line with respect to
the weight

\begin{equation}
w^{\left( 2m\right) }\left( x,a\right) =\frac{x^{2\left( a-1\right) }}{%
\left( \mathit{L}_{2m}^{\left( 1-2a\right) }(-\gamma x/\left\vert
a_{-(2m+1)}\right\vert )\right) ^{2}}.
\end{equation}

In the case $(i)$, the situation is quite different since the ground state
of $V^{\left( n\right) }$ is associated to the RS function $w_{0}^{\left(
n\right) }(x;a)$ and the superpartner of the extended potential $V^{\left(
n\right) }(x;a)$ is now given by

\begin{equation}
\widetilde{V}^{\left( n\right) }(x;a)=V^{\left( n\right)
}(x;a)+2w_{0}^{\left( n\right) \prime }(x;a),\ n\geq 0,
\label{SUSYpartextKC}
\end{equation}%
as for the $L1$ and $L2$ extensions of the isotonic oscillator \cite%
{grandati3} but in the ERKC case $V^{\left( n\right) }$ does not inherit of
the shape invariance properties of the initial TSIP.

The physical eigenstates for the energies $E_{k}\left( a\right) ,\ k\geq 0$
of $\widehat{H}^{\left( n\right) }(a)$ are given by

\begin{equation}
\psi _{k}^{\left( n\right) }(x;a,\gamma )\sim x^{a-1}e^{-\gamma x/2a_{k}}%
\frac{N_{a,k}^{(n)}\left( x\right) }{\mathit{L}_{n}^{\left( 1-2a\right)
}(\gamma x/\left\vert a_{-(n+1)}\right\vert )},
\end{equation}%
and the function\bigskip s 
\begin{equation}
C_{k}^{\left( n\right) }\left( x,a\right) =e^{-\gamma
x/2a_{k}}N_{a,k}^{(n)}\left( x\right) ,\ k\geq 0,
\end{equation}%
are orthogonal on the positive half line with respect to the weight

\begin{equation}
w^{\left( n\right) }\left( x,a\right) =\frac{x^{2\left( a-1\right) }}{\left( 
\mathit{L}_{n}^{\left( 1-2a\right) }(\gamma x/\left\vert
a_{-(n+1)}\right\vert )\right) ^{2}}.
\end{equation}

\section{Conclusion}

In this paper we have shown that the method previously developed for the
isotonic potential \cite{grandati3}, can be used to generate in a direct and
systematic way the solvable regular rational extensions for all the
exceptional first category TSIP. This approach is based on DBT
transformations built from excited states RS functions regularized via the
use of discrete symmetries of the initial potential.

The results are quite different from those obtained for the isotonic
oscillator (which is the unique exceptional second category TSIP). Each
exceptional first category TSIP admits only one series of regular rational
extensions. Generally, as for the $L3$ series of rational extensions of the
isotonic potential, it can be obtained only from regularized excited states
associated to even quantum numbers and the DBT can be viewed as a backward
SUSY partnership. The isospectrality is not strict and the spectrum of the
extended potential presents a supplementary lower level. The ERKC potential
constitutes an exception since extended potentials can be also obtained from
regularized excited states RS functions associated to odd quantum numbers
for some range of values of the "angular momentum" parameter $a$. They are
in this case strictly isospectral to the original potential.

\section{Acknowledgments}

I would like to thank A.\ B\'{e}rard, R.\ Milson and C.\ Quesne for
stimulating discussions and very interesting suggestions.

\end{document}